\documentclass[aps,prd,twocolumn,showpacs,notitlepage,eqsecnum,
superscriptaddress,nofootinbib]{revtex4-1}

\usepackage[centertags]{amsmath} \usepackage{amssymb} \usepackage{latexsym}
\usepackage{enumerate} \usepackage{graphicx} \usepackage{mathrsfs}
\usepackage[colorlinks]{hyperref} \usepackage{stmaryrd} \usepackage{import}
\usepackage{tensor} \usepackage[usenames,dvipsnames]{xcolor} \usepackage{bm}
\usepackage{multirow} \usepackage{breakurl} \usepackage{float}
\usepackage{verbatim} \usepackage{mathrsfs}

\allowdisplaybreaks[1]

\definecolor{CiteColor}{rgb}{0,0.5,0} \hypersetup{citecolor=CiteColor}
\definecolor{RefColor}{rgb}{0.55,0,0} \hypersetup{linkcolor=RefColor}
\definecolor{darkgreen}{rgb}{0.2,0.7,0.2}

\begin{document}

\title{Highly eccentric EMRI waveforms via fast self-forced inspirals}

\author{Jonathan McCart}
\email{jmccart6@gatech.edu}
\affiliation{Department of Physics and Astronomy, 
State University of New York at Geneseo, New York 14454, USA}
\affiliation{Machine Learning Center, Georgia Institute of Technology, Atlanta, Georgia 30332, USA}
\author{Thomas Osburn}
\email{tosburn@geneseo.edu}
\affiliation{Department of Physics and Astronomy, 
State University of New York at Geneseo, New York 14454, USA}
\affiliation{Department of Physics and Astronomy, 
University of North Carolina, Chapel Hill, North Carolina 27599, USA}
\author{Justin Y. J. Burton}
\affiliation{Department of Physics, 
Emory University, Atlanta, Georgia 30322, USA}
\affiliation{Department of Physics and Astronomy, 
State University of New York at Geneseo, New York 14454, USA}

\begin{abstract}

We present new developments and comparisons of competing inspiral and waveform models for highly eccentric non-spinning extreme and intermediate mass-ratio inspirals (EMRIs and IMRIs). Starting from our high eccentricity self-force library, we apply the near-identity transform (NIT) technique to rapidly compute highly eccentric self-forced inspirals for the first time. Upon evaluating our approximate NIT results via comparison with full self-force inspirals, we couple our accurate and streamlined inspiral data to potential waveform generation schemes. We find that, although high eccentricity strains the NIT method, NIT inspirals are consistent with full self-force inspirals for EMRIs. However, our NIT implementation (at 1\textsuperscript{st} post-adiabatic order) is not able to achieve LISA-motivated accuracy goals for highly eccentric IMRIs. Our most sophisticated waveforms are devised through a new technique that efficiently connects NIT orbital parameters to Teukolsky amplitudes and phases. We compare these sophisticated Teukolsky waveforms to those with synthesized (summing over harmonics) amplitudes based on a kludge. We find that, assuming identical worldlines (so that dephasing is negligible), kludge waveforms compare favorably to Teukolsky waveforms for non-spinning bodies.

\end{abstract}

\maketitle

\section{Introduction}

As ground-based gravitational wave observations continue to reveal rich features of compact binary systems and their astrophysical environment~\cite{Abbott_2019}, further direction of attention towards future generations of gravitational wave detectors is increasingly important. One such detector is the Laser Interferometer Space Antenna (LISA)~\cite{baker2019laser}, which will probe lower frequency gravitational waves for the first time. Exploring this uncharted band of the gravitational wave spectrum is expected to uncover exciting new sources such as super-massive black hole binaries and extreme mass-ratio inspirals (EMRIs)~\cite{Amaro_Seoane_2018}. EMRIs consist of a small compact body (with a typical mass $\mu\sim 10 M_\odot$) orbiting a massive black hole (with a typical mass $M\sim 10^6 M_\odot$). EMRIs are especially valuable systems to investigate because of their unique features, such as high eccentricity~\cite{HopmAlex05}. Additionally, the long durations of EMRIs will facilitate precise strong-field tests of general relativity~\cite{Barack_2007,berry2019unique}. Another type of LISA source related to EMRIs is intermediate mass-ratio inspirals (IMRIs), which involve compact bodies that are slightly closer in mass (such as $\mu\sim 100 M_\odot$ and $M\sim 10^5 M_\odot$). This work applies black hole perturbation theory to investigate and improve gravitational waveform models for highly eccentric EMRIs and IMRIs.

When the mass-ratio, $\epsilon \equiv \mu/M$, is sufficiently small, black hole perturbation theory is a practical and accurate approach to the two-body problem of general relativity. Other potential approaches, such as post-Newtonian theory or numerical relativity, are less suitable to access important EMRI features like relativistic velocities or disparate masses (black hole perturbation theory accommodates both of those). This perturbative approach involves an expansion of the gravitational field in powers of the small mass-ratio. The leading term in the expansion represents the massive stationary black hole, while higher order terms are perturbations that characterize two-body interactions. In accordance with black hole perturbation theory, orbital dynamics are influenced by the self-force~\cite{Mino_1997,Quinn_1997} acting on the small body. Through these processes, both waveform and orbital features are accessible by quantifying the behavior of gravitational perturbations.

In pursuit of waveform templates that meet or exceed LISA requirements, it is critical to consider how inspiraling orbit calculations affect the accuracy of EMRI models. Phase accuracy requirements in particular motivate the level of sophistication necessary when approximating the worldline. In order to facilitate optimal LISA detections (and subsequent parameter estimation), the phase error must be significantly smaller than $\sim 1$ radian. Therefore, when analyzing how the small mass-ratio ($\epsilon$) influences the equations of motion, we seek a sufficiently high order of expansion to ensure the orbital phases are accurate to $\mathcal{O}(\epsilon^0)$ at minimum (it is plausible that higher order terms could have large enough coefficients to also necessitate $\mathcal{O}(\epsilon^{1})$ contributions, see Sec.~\ref{sec:FullVsNIT} where we measure the size of certain higher order coefficients for the first time). This phase expansion resides within the post-adiabatic framework~\cite{HindFlan08}, and the minimum expansion order for sufficient phase accuracy (through $\mathcal{O}(\epsilon^0)$) is the 1\textsuperscript{st} post-adiabatic order.

The mechanism that drives orbital evolution and decay (with associated accumulated phases) is the gravitational self-force~\cite{Mino_1997,Quinn_1997}. It is well known that, to achieve minimum phase requirements (accurate through 1st post-adibatic order), the conservative part of the self-force must be expanded through 1\textsuperscript{st} order while the dissipative part of the self-force must be expanded through 2\textsuperscript{nd} order~\cite{HindFlan08}. Although recent breakthroughs have begun to access the crucial 2\textsuperscript{nd} order self-force~\cite{PounETC20,warburton2021gravitationalwave}, we do not include 2\textsuperscript{nd} order effects here. Rather, our goal is to apply 1\textsuperscript{st} order (conservative and dissipative) self force results in a framework that is flexible to incorporate 2\textsuperscript{nd} order effects as those calculations mature. Note that the aforementioned phase accuracy requirements preclude (for effective LISA data analysis) usage of other relaxed inspiral approximations including kludge inspirals, post-Newtonian inspirals, or adiabatic inspirals (each of those would introduce a phase error of $\sim 1$ radian or greater unless they also incorporate self-force information through a strategy such as effective one-body theory~\cite{Buonanno_1999,Antonelli_2020}). One unfortunate disadvantage of self-forced inspiral calculations is the potential for considerable computational expense. A straightforward implementation would require many numerical integration steps during each orbital oscillation (and for $\sim \epsilon^{-1}$ total oscillations). Prior work has addressed this challenge by applying a near-identity transform (NIT)~\cite{KevoCole96} to eliminate oscillatory behavior in the equations of motion~\cite{VandWarb18}; here we present new enhancements that achieve NIT applicability for a wider range of configurations (see Sec.~\ref{sec:FullVsNIT} for results demonstrating those enhancements).

In addition to theoretical considerations (expansion orders, etc.), it is similarly vital to ensure that inspiral models exhibit the astrophysical features expected of realistic EMRIs. One important new feature that we incorporate into the inspiral model presented here is highly eccentric orbital motion. EMRIs can have eccentricities up to $e\simeq 0.75$~\cite{HopmAlex05}, and here we achieve fast (NIT) self-forced inspirals in that regime for the first time. In addition to eccentricity, there are other expected astrophysical features that will be left to future work, such as spin for the larger~\cite{WarbBara11,Vand16,Vand18,NasiOsbu19,nasipak2021resonant} and/or smaller~\cite{BurkKhan15,WarbOsbu17,Piovano_2020,Ruangsri_2016} binary components. Another astrophysical consideration regarding EMRIs is that the gravitational waves from these two-body systems may~\cite{BongHuan19} or may not~\cite{BaraCard14} be influenced by environmental effects (perhaps on a case-by-case basis), and in this work we do not consider possible influences from EMRI environments.

Our discussion thus far has concerned waveform phases, but we could also consider how inspiral calculations affect tolerances for waveform amplitudes. Generally, if the worldline is calculated at sufficient accuracy to achieve phase requirements, then that orbital description would not be a limiting factor for accurate waveform amplitudes (see Sec.~\ref{sec:waveforms}). Rather, the specific procedure for generating waveforms (by post-processing the worldline) controls amplitude accuracy. While black hole perturbation theory and the self-force are critical in achieving waveform (via inspiral) phase accuracy requirements, perhaps certain approximation orders could be relaxed in determining waveform amplitudes during inspiral evolution (maybe some sort of kludge could suffice). Accordingly, one outcome of this work is to assess the relative accuracy (and relative LISA-related outcomes) for calculating amplitudes from competing waveform generation schemes (kludge vs. Teukolsky, see Sec.~\ref{sec:TeukVsKludge}) when applying our newly enhanced (see Sec.~\ref{sec:nit}) inspiral model. Ultimately, our model and analysis is approximately equivalent to an extension of FastEMRIWaveforms~\cite{Chua_2021,katz2021fastemriwaveforms} results to include vital conservative self-force effects, except without certain algorithmic enhancements (we discuss those potential enhancements further in Sec.~\ref{sec:CandFD}).

\section{Self-forced inspirals}
\label{sec:inspirals}

\subsection{Self-force calculations}

The 1\textsuperscript{st} order self-force is calculated from a small (1\textsuperscript{st} order in $\epsilon$) metric perturbation, $h_{\mu\nu}$, that describes how gravitational fields surrounding the central black hole are influenced by two-body interactions. Then, the Einstein equations are expanded in powers of $\epsilon$ through 1\textsuperscript{st} order
\begin{align}
&G_{\alpha\beta}(g_{\mu\nu}+h_{\mu\nu} + \mathcal{O}(\epsilon^2) )= T_{\alpha\beta},
\end{align}
where $G_{\alpha\beta}$ is the Einstein tensor, $g_{\mu\nu}$ is the stationary metric of the central black hole (Schwarzschild spacetime for this work), and $T_{\alpha\beta}$ is the stress-energy tensor. Upon selecting an appropriate gauge (we adopt the Lorenz gauge), we then solve the field equations for $h_{\mu\nu}$ via separation of variables
\begin{align}
\label{eq:sep}
&h_{\mu\nu}(t, r, \theta, \phi) = \sum_{lm\omega} \sum_{k=1}^{10} R^{(k)}_{lm\omega}(r) \; Y^{(k)lm}_{\mu\nu}(\theta,\phi) \; e^{-i\omega t} ,
\end{align}
where $R^{(k)}_{lm\omega}$ are radial separation functions and $Y^{(k)lm}_{\mu\nu}$ are the tensor spherical harmonics. Notice that we have adopted a frequency-domain approach. To calculate the self-force, the metric perturbation is subjected to a regularization procedure that carefully ameliorates the Coulomb-like divergence near the small body's worldline (we adopt the mode-sum regularization procedure~\cite{BaraOri00,Bara01,BaraOri03}). Improving upon pre-existing efforts to calculate the self-force for smaller eccentricities~\cite{BaraSago10,AkcaWarb13}, we enact the above logic through our recently developed high eccentricity self-force code~\cite{OsbuFors14,HoppFors15}. The inputs for our code, $e$ (eccentricity) and $p$ (semi-latus rectum $\sim$ separation/$M$), describe the size and shape of the orbit. Our self-force code represents the past worldline as a geodesic, which is an approximation that may benefit from future refinement. 

By executing our code in parallel, we have calculated the self-force for $\sim$44000 unique pairs of orbital parameters ($e\lesssim 0.8$, $p\lesssim 50$) at a cost of $\sim$5000 CPU hours. During inspiral evolution we also require self-force values from orbital parameters in between those that were pre-computed, so we have implemented a scheme to interpolate across $e$ and $p$. Inherent challenges inhibited the generation of results on a rectangular grid (the code avoids certain quantifiable orbital configurations~\cite{OsbuFors14}) so we perform least-squares fitting to determine a patchwork of local interpolation coefficients (with overlapping interpolation domains). The resulting interpolants describe the Fourier coefficients for each component of the self-force. 

\subsection{Rapid inspiral calculations: NIT}
\label{sec:nit}

It is relatively straightforward to harness self-force results for long-term inspiral evolutions. One powerful technique is to re-parametrize the equations of motion in terms of evolving orbital elements ($p$, $e$, etc.) that describe a sequence of geodesics tangent to the inspiraling worldline~\cite{PounPois08,GairFlan11}. Note that this orbital re-parametrization via osculating elements is not synonymous with the usage of geodesics to represent the past worldline during self-force calculations (although we do both in this work). It is convenient to characterize the Schwarzschild orbital radius, $r_p$, in terms of the Darwin phase, $v$~\cite{Darw59,Darw61}
\begin{align}
\label{eq:dar}
r_p(p,e,v) = \frac{pM}{1+e \cos{v}} ,
\end{align}
which implies geometric definitions for $p$ and $e$ based on the maximum and minimum values of $r_p$. Incorporating the Schwarzschild orbital azimuth, $\phi_p$, and time, $t$, assembles a complete set of variables ($v$, $p$, $e$, $\phi_p$, $t$) to track the inspiral evolution.

To generate equations of motion, it is convenient to introduce a new time parameter that is similar to (but not entirely equivalent to) $v$. This new time parameter, $\chi$, is designed so that the difference $v-\chi$ follows the slowly changing initial Darwin phase ($v$ at $t=0$) for a tangent geodesic (not to be confused with $v_0$ for the inspiraling worldline). Then, the equations of motion involve derivatives with respect to $\chi$ 

\begin{align}
\label{eq:EOMi}
& \frac{dv}{d\chi} = 1 + \epsilon \, f_v(p,e,v) ,
\\
& \frac{dp}{d\chi} = \epsilon \, \mathcal{F}_p(p,e,v) ,
\\
& \frac{de}{d\chi} = \epsilon \, \mathcal{F}_e(p,e,v) ,
\\
&\frac{d\phi_p}{d\chi} = f_\phi(p,e,v) ,
\\
&\frac{dt}{d\chi} = f_t(p,e,v) ,
\label{eq:EOMf}
\end{align}
where $f_v$, $\mathcal{F}_p$, and $\mathcal{F}_e$ depend on the self-force, and their forms (along with $f_\phi$ and $f_t$) are given in~\cite{VandWarb18}. This strategy has been implemented successfully in past work~\cite{WarbAkca12,OsbuWarb16,WarbOsbu17}, but these equations of motion have unfortunate properties that hinder computational efficiency. Specifically, it is the explicit $v$ dependence in the rates-of-change that causes solutions to oscillate rapidly (during each orbital period). These oscillations are present because $v$, during its monotonic growth, appears within sines and cosines in Eqs.~\eqref{eq:EOMi}-\eqref{eq:EOMf}. Therefore, numerical integration efforts require many integration steps during each oscillation (and for $\sim \epsilon^{-1}$ total oscillations). In contrast, if the equations of motion did not rely explicitly on phases (such as $v$), only a modest number of integration steps would be necessary (the amount would be independent of $\epsilon$ rather than inversely proportional). However, the existing equations of motion have been motivated by appropriate physics, so great care must be taken when seeking a compatible transformation that avoids explicit $v$ dependence.

To conquer this challenge, we follow~\cite{VandWarb18} and perform an averaging transformation whose inverse should successfully approximate the original (oscillating) orbital variables within required tolerances. This process for generating compatible equations of motion without rapid oscillations is called a near-identity transformation (NIT)~\cite{KevoCole96}. We distinguish NIT (non-oscillating) variables from the original orbital parameters by placing a tilde ($\sim$) above NIT variables: ($\tilde{v}$, $\tilde{p}$, $\tilde{e}$, $\tilde{\phi_p}$, $\tilde{t}$\,). Then our goal is to, starting from the set ($v$, $p$, $e$, $\phi_p$, $t$), find a specific averaging transformation that achieves the following inverse NIT accuracy requirement
\begin{align}
\label{eq:NITaccuracy}
&\xrightarrow{\;\mathrm{NIT}\;} (\tilde{v}, \tilde{p}, \tilde{e}, \tilde{\phi_p}, \tilde{t}\,) \xrightarrow{\mathrm{inverse}} (v, p, e, \phi_p, t) + \mathcal{O}(\epsilon) ,
\end{align}
although, in practice $\tilde{v}$, $\tilde{p}$, and $\tilde{e}$ are inherently within $\mathcal{O}(\epsilon)$ of $v$, $p$, and $e$, so the inverse transformation will be necessary for some variables (like $\phi_p$ and $t$) but may not be critical for all variables. 

Precise details of the transformation that simultaneously eliminates oscillations while maintaining consistency with Eq.~\eqref{eq:NITaccuracy} are provided in~\cite{VandWarb18}, but we summarize a few key features here. The NIT equations of motion have the following form:
\begin{align}
\label{eq:NITi}
& \frac{d\tilde{v}}{d\chi} = 1 + \epsilon \, \tilde{f}_v^{(1)}(\tilde{p},\tilde{e}) ,
\\
& \frac{d\tilde{p}}{d\chi} = \epsilon \, \tilde{\mathcal{F}}_p^{(1)}(\tilde{p},\tilde{e}) + \epsilon^2 \, \tilde{\mathcal{F}}_p^{(2)}(\tilde{p},\tilde{e}) ,
\\
& \frac{d\tilde{e}}{d\chi} = \epsilon \, \tilde{\mathcal{F}}_e^{(1)}(\tilde{p},\tilde{e}) + \epsilon^2 \, \tilde{\mathcal{F}}_e^{(2)}(\tilde{p},\tilde{e}) ,
\\
&\frac{d\tilde{\phi_p}}{d\chi} = \tilde{f}_\phi^{(0)}(\tilde{p},\tilde{e}) + \epsilon \, \tilde{f}_\phi^{(1)}(\tilde{p},\tilde{e}) ,
\\
&\frac{d\tilde{t}}{d\chi} = \tilde{f}_t^{(0)}(\tilde{p},\tilde{e}) + \epsilon \, \tilde{f}_t^{(1)}(\tilde{p},\tilde{e}) .
\label{eq:NITf}
\end{align}
Notice that neither $v$ nor $\tilde{v}$ appear in the right-hand side of the NIT equations of motion, which successfully avoids costly oscillations. Terms with a $(0)$ superscript do not involve the self-force, terms with a $(1)$ superscript are averages of products involving the self-force, and terms with a $(2)$ superscript are averages of products involving derivatives of the self-force with respect to $p$ and $e$ (those final terms will, when available, also include the averaged 2\textsuperscript{nd} order self-force). Numerical integration of Eqs.~\eqref{eq:NITi}-\eqref{eq:NITf} is rapid, and the fast computation time has negligible dependence on $\epsilon$. Inversion of $\tilde{t}$ and $\tilde{\phi}_p$ occurs by adding oscillatory functions based on geodesic characteristics
\begin{align}
    t &= \tilde{t} - Z^{(0)}_t(\tilde{p},\tilde{e},\tilde{v}) + \mathcal{O}(\epsilon) ,
    \\
    \phi_p &= \tilde{\phi}_p - Z^{(0)}_\phi(\tilde{p},\tilde{e},\tilde{v})+ \mathcal{O}(\epsilon) .
    \label{eq:phitilde}
\end{align}
Derivations of $Z^{(0)}_t$ and $Z^{(0)}_\phi$ are available in~\cite{VandWarb18}, and we provide an expression for $Z^{(0)}_\phi$ here as an example
\begin{align}
    & Z^{(0)}_\phi = \sqrt{\frac{4\tilde{p}}{\tilde{p}+2\tilde{e}-6}}\Bigg( \Big( \frac{\tilde{v}-\pi}{\pi}\Big) \, K\Big( \frac{4\tilde{e}}{\tilde{p}+2\tilde{e}-6} \Big) \notag
    \\
    &\qquad\qquad\qquad\qquad\;\;\, - F\Big(\frac{\tilde{v}-\pi}{2} \Big| \frac{4\tilde{e}}{\tilde{p}+2\tilde{e}-6} \Big) \Bigg)\, ,
\end{align}
where $K$ and $F$ are elliptic integrals following the conventions of Mathematica.

One necessary development for NIT implementation involves reconciling our self-force interpolation scheme with the need to calculate derivatives of the self-force with respect to $p$ and $e$. Because we rely on a patchwork of local interpolations, there is a small level of non-smoothness when transitioning from one interpolation domain to the next. Unfortunately, this lack of perfect smoothness is amplified during differentiation with respect to $p$ and $e$. To resolve this issue, we leverage a helpful feature of our interpolation scheme: overlapping adjacent interpolation domains. By using smooth transition functions in a weighted average of adjacent domains, we are able to implement a perfectly smooth interpolation scheme that is consistent with the original tolerances. Upon achieving a smooth model that covers the appropriate parameter space, we re-sampled the values on a rectangular grid in $e$ vs. $p-2e$. Our NIT inspiral computations follow from interfacing our high eccentricity self-force library with a modified version of the Fast Self-Forced Inspirals package~\cite{VandWarb18} of the Black Hole Perturbation Toolkit~\cite{toolkit}. Modifications to accommodate high eccentricity include increasing the number of harmonics used during Fourier decomposition, increasing the maximum $p$ to 26, and increasing the maximum $e$ to 0.8; although, we consider usage above $e\simeq 0.75$ to be experimental because our frequency domain self-force code struggles when $e$ approaches 1.

\begin{figure}
\includegraphics[width=3.3in]{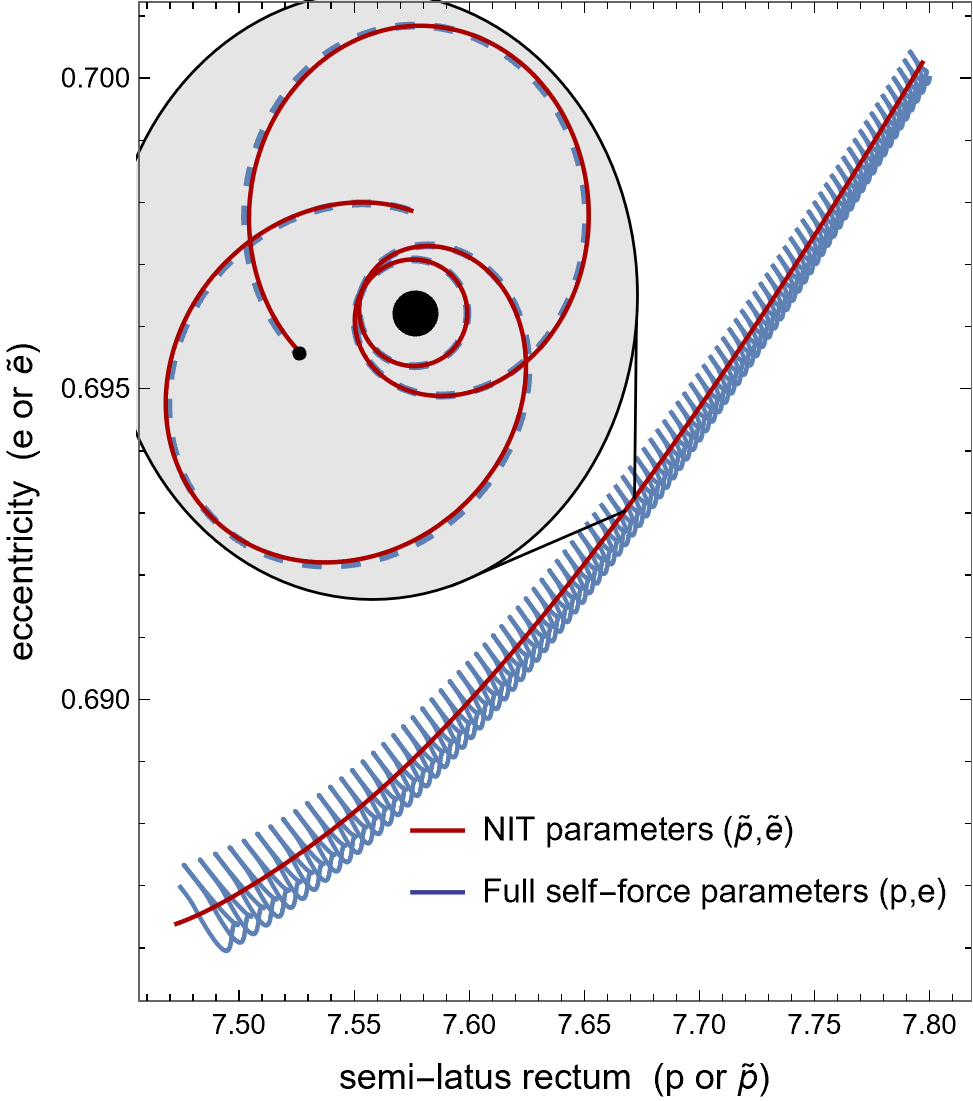}
\caption{\label{fig:orbit} Eccentricity vs. semi-latus rectum is plotted for a NIT inspiral and a full self-force inspiral with equivalent initial parameters. Initial matching occurs by ensuring that $\tilde{p}_0$ and $\tilde{e}_0$ are consistent with $p_0$ and $e_0$ after applying the inverse NIT (this guarantees matched initial frequencies). After initial matching it is not essential to invert $\tilde{p}$, $\tilde{e}$, and $\tilde{v}$ because they are within $\mathcal{O}(\epsilon)$ of $p$, $e$, and $v$ (which is sufficient for our applications). However, it is typically necessary to invert $\tilde{t}$ and $\tilde{\phi}_p$ (to recover $t$ and $\phi_p$). For illustrative purposes the parameters $p_0=7.8$, $e_0=0.7$, and $\epsilon=5\times 10^{-5}$ are depicted (comparisons with more demanding parameters are discussed starting in Sec.~\ref{sec:TeukWave}). Notice that the NIT parameters do not oscillate (by design), which is why that method is ``fast" (and the full self-force method is ``slow"). The oval inset depicts a mid-evolution reconstruction of the two trajectories in polar coordinates with $[r,\phi]^\text{NIT} = [r_p(\tilde{p},\tilde{e},\tilde{v}),\,\tilde{\phi}_p-Z^{(0)}_\phi(\tilde{p},\tilde{e},\tilde{v})]$ (solid red) and $[r,\phi]^\text{Full} = [r_p(p,e,v),\,\phi_p]$ (dashed blue). }
\end{figure}

It is important to verify our enhanced high eccentricity NIT implementation. It has been shown that the approximate inverse transformation does not adversely affect EMRI phase accuracy at low eccentricities~\cite{VandWarb18}, and here we are able to make similar assessments at high eccentricities. Figure~\ref{fig:orbit} illustrates high eccentricity results from both our fast NIT model and the prior full self-force model based on original orbital variables (Fig.~\ref{fig:orbit} parameters were chosen to highlight orbital features rather than stress-test the NIT method). In Sec.~\ref{sec:TeukWave} we demonstrate long-duration agreement between highly eccentric NIT and full self-force EMRI evolutions (see Fig.~\ref{fig:FullVsNIT}) where the full self-force inspiral consumed $\sim 173$~seconds of computing time while the NIT method consumed only $~\sim 0.0401$~seconds of computing time ($\sim 4000$ times faster). However, in the intermediate mass-ratio regime the accuracy of the NIT method (through 1\textsuperscript{st} post-adiabatic order) deteriorates depending on eccentricity (see Sec.~\ref{sec:FullVsNIT}).

\section{Waveforms}
\label{sec:waveforms}

\subsection{Kludge Waveforms}
\label{sec:KludgeWave}

Equipped with two models to compute self-forced worldlines (slow full self-force vs. fast but approximate NIT), we now investigate possible strategies for post-processing orbital data to generate waveforms. All waveform schemes share a common angular description in terms of spin-weighted spherical harmonics
\begin{align}
\label{eq:hwave}
&h_+ - i\, h_\times = \frac{1}{r}\sum_{lm} H_{lm}(t_\text{ret})\;_{-2}Y_{lm}(\theta, \phi) ,
\end{align}
where $t_\text{ret}\equiv t-r$ (an approximate far-zone expansion of the true retarded time). Therefore, competing waveform schemes differ in their representation of $H_{lm}$. Before exploring procedures to generate authentic relativistic waveforms via numerical Teukolsky solutions, we first investigate less sophisticated (but perhaps more accessible) approaches. 

One common scheme follows from a weak-field approximation where the (in this context) relativistic worldline is artificially mapped to Minkowski spacetime (typically the Schwarzschild $r$ coordinate is interpreted as a Euclidean distance). This approximation facilitates so-called``kludge" waveforms by attaching a point source to that artificial worldline and solving the flat spacetime wave equation. Upon appropriate far-zone extraction of $h_+$ and $h_\times$, $H_{lm}^\text{kludge}$ can be identified for each multipole. As an example, we report the $(l,m)=(2,2)$ mode of $H_{lm}^\text{kludge}$ for our case involving eccentric equatorial motion (see Appendix~\ref{sec:kludge_app} for the derivation)
\begin{align}
&H_{2,2}^\text{kludge}(t_\text{ret}) = 4\,\mu\,\sqrt{\frac{\pi}{5}} \,e^{-2i\phi_p} \bigg(\dot{r}_p^2-4\,i\,r_p\,\dot{r}_p\,\dot{\phi}_p \notag
\\&\qquad\qquad\qquad\qquad +r_p\left( \ddot{r}_p-r_p(2\,\dot{\phi}_p^2+i\,\ddot{\phi}_p) \right) \bigg) ,
\label{eq:kludge}
\end{align}
where instantaneous orbital quantities are evaluated at $t_\text{ret}$ and dots signify derivatives with respect to $t_\text{ret}$. It is straightforward to use tangent geodesic properties to express $r_p$, $\dot{r}_p$, $\ddot{r}_p$, $\dot{\phi}_p$, and $\ddot{\phi}_p$ in terms of $v$, $p$, and $e$, which are available (along with $\phi_p$) from the inspiral calculation. Truncating the multipole expansion after a small number of terms would also introduce a slow speed approximation, but in principle an arbitrary number of multipoles can be incorporated. It is not entirely necessary to express kludge waveforms using a series of spin-weighted spherical harmonics (as in Eq.~\eqref{eq:hwave}), but this decomposition is helpful in our comparison of kludge waveforms vs. Teukolsky waveforms to assess performance regarding LISA data analysis.

\subsection{Evolving Teukolsky Snapshots and the NIT}
\label{sec:TeukWave}

We use the term ``Teukolsky snapshot" in reference to the periodic waveform associated with a geodesic Teukolsky source during a narrow time window. Although actual EMRI waveforms are not entirely periodic (due to the slow inspiral evolution), the true waveform at a certain time is closely matched with the snapshot from an associated tangent geodesic. Carefully quantifying these periodic Teukolsky snapshots is an important a step towards our eventual goal of generating accurate waveforms from entire inspirals. 

In the absence of radiation-reaction, bound-stable orbital motion around a black hole produces a waveform with the same discrete frequencies as its source. For eccentric-equatorial orbits, the spectrum of gravitational wave frequencies, according to the above logic, is governed by the two fundamental orbital frequencies, $\Omega_\phi$ and $\Omega_r$
\begin{align}
\label{eq:Hperiodic}
& H^\text{Teuk}_{lm}(t_\text{ret}) = \sum_n A_{lmn} \; e^{-i(m\Omega_\phi + n \Omega_r)t_\text{ret}} ,
\end{align}
where the complex amplitudes, $A_{lmn}$, are related to non-homogeneous Teukolsky amplitudes~\cite{WarbOsbu17}.  Moving forward, we introduce slightly modified terminology by making the notationally abusive replacement $t_\text{ret} \rightarrow t$. We find that a number of harmonics totaling $|n_\text{max}| \simeq 40$ is sufficient to faithfully represent our highest eccentricity cases ($e\simeq0.75$). Apparent similarities between Eq.~\eqref{eq:sep} and the combined Eqs.~\eqref{eq:hwave} and~\eqref{eq:Hperiodic} are no coincidence, indeed, $A_{lmn}$ is related to the asymptotic ($r\simeq \infty$) behavior of $R_{lm\omega}^{(k)}$. The orbital frequencies are characterized based on periodicity observed during successive periastrons
\begin{align}
\label{eq:Omega_phi}
&\Omega_\phi^\text{geodesic} = \frac{\phi^{(j+1)}_p-\phi^{(j)}_p}{t^{(j+1)}-t^{(j)}} ,
\\
\label{eq:Omega_r}
&\Omega_r^\text{geodesic} = \frac{2\pi}{t^{(j+1)}-t^{(j)}} ,
\end{align}
where $(j)$ superscripts represent an integer used to count periastron passages. Ignoring radiation-reaction, these fundamental frequencies would be unchanged from one periastron to the next.

This clean snapshot picture based on discrete frequencies becomes muddled during inclusion of self-force effects. Because the self-force involves dissipation, the source (and, therefore, the associated inhomogeneous waveform) will exhibit a continuous frequency spectrum. However, the slow inspiral evolution does involve a waveform spectrum that is sharply peaked surrounding quasi-periodic harmonics (when sampled for small durations). One possible strategy to access these quasi-periodic waveforms without introducing any additional approximations is to input the entire inspiral as a source to drive the time-domain field equations. There are two unfortunate drawbacks of this strategy. First, because the inspiral duration is very long ($t_\text{max} \sim M \epsilon^{-1}$) such calculations would be (arguably) impractical from the viewpoint of computational expense. Second, for accurate inspirals informed by the self-force, the relevant field equations would have already been solved during self-force calculations (so it would seem redundant to solve them again during waveform generation).

Instead, we will adopt an attractive alternative involving a convenient approximation. Our strategy approximates the waveform as a continuous sequence of geodesic Teukolsky snapshots. This ``evolving snapshots" technique was implemented in~\cite{WarbOsbu17}, and more recently has been referred to as a ``multivoice" decomposition~\cite{Hughes_2021}. While this representation of the past worldline is not exact, by using the accumulated orbital phases to accurately inform the waveform phase we should avoid all but a few negligible amplitude errors. As a side note, while tangent geodesics are sufficient for determining accurate waveform amplitudes, accurate self-force calculations may require comparatively improved approximations of the past worldline. To implement evolving snapshots, we first pre-compute Teukolsky amplitudes ($A_{lmn}$) for all necessary modes ($l$, $m$, and $n$) and also for a dense set of orbital parameters ($p$ and $e$). In principle, these Teukolsky amplitudes and the self-force can be pre-computed simultaneously with minimal additional effort (although they were computed separately here). Then we interpolate across different orbital configurations (similar to our approach for the self-force) so that each Teukolsky amplitude is represented as a function of $p$ and $e$. Under this scheme, Eq.~\eqref{eq:Hperiodic} is replaced with the following:
\begin{align}
\label{eq:Hcont}
& H^\text{Teuk}_{lm}(t) = \sum_n A_{lmn}(p,e) \; e^{-i \Phi_{mn}} ,
\end{align}
where $p$, $e$, and each frequency's accumulated waveform phase, $\Phi_{mn}$, depend on $t$. To determine $\Phi_{mn}$, we consider how waveform phases are affected by time-dependent orbital frequencies
\begin{align}
\label{eq:phase_int}
&\Phi_{mn}(t) = m \int_0^t \Omega_\phi \, dt + n \int_0^t \Omega_r \, dt.
\end{align}
By generalizing $\Omega_\phi$ and $\Omega_r$ to embody quasi-periodic scenarios, these integrals capture the effect of slowly evolving frequencies on the accumulated waveform phases. However, great care must be taken to define $\Omega_\phi$ and $\Omega_r$ appropriately so that the waveform phases are consistent with the accumulated orbital phases. Specifically, the following kinematic constraints should be enforced:
\begin{align}
\label{eq:phiPhase}
&\int_{t^{(j)}}^{t^{(j+1)}} \Omega_\phi \, dt = \phi^{(j+1)}_p-\phi^{(j)}_p ,
\\
\label{eq:rPhase}
&\int_{t^{(j)}}^{t^{(j+1)}} \Omega_r \, dt = 2 \pi ,
\end{align}
(recall that $(j)$ superscripts count successive periastrons). Equations~\eqref{eq:phiPhase} and~\eqref{eq:rPhase} can be conceptualized as generalizations of Eqs~\eqref{eq:Omega_phi} and \eqref{eq:Omega_r} for quasi-periodic motion. For strategies that rely on direct integration of Eq.~\eqref{eq:phase_int}, we recommend using Eqs.~\eqref{eq:phiPhase} and \eqref{eq:rPhase} as a benchmark for consistency between the waveform phases and orbital phases (although, we present here an alternative strategy that does not require additional integrals).

To illustrate the importance and validity of Eqs.~\eqref{eq:phiPhase} and \eqref{eq:rPhase}, consider an inspiral evolution ending at the $j^\text{th}$ periastron (with $t^{(0)} = \phi^{(0)}_p = v^{(0)} = 0$ throughout for simplicity). According to Eqs.~\eqref{eq:phase_int}-\eqref{eq:rPhase}, the accumulated waveform phases at $t=t^{(j)}$ are
\begin{align}
\label{eq:periPhase}
&\Phi_{mn}^{(j)} = m\, \phi^{(j)}_p + n\, v^{(j)} ,
\end{align}
(recall that $v$ is defined as accumulating $2\pi$ radians per periastron: $v^{(j)} = 2\pi j$). The instantaneous waveform value at the $j^\text{th}$ periastron is then governed by Eq.~\eqref{eq:Hcont}
\begin{align}
H^\text{Teuk}_{lm}(t^{(j)}) &= \sum_n A_{lmn} \; e^{-i \Phi^{(j)}_{mn}} \notag
\\ &= \sum_n A_{lmn} \; e^{-i m \phi^{(j)}_p} e^{-i n v^{(j)}} \notag
\\ &= \sum_n \left[ A_{lmn} \; e^{-i m \phi^{(j)}_p}\right] ,
\label{eq:sqBrack}
\end{align}
where $e^{-i n v^{(j)}}=1$ because $n$ and $j$ are integers. Notice that the factors inside the square brackets in Eq.~\eqref{eq:sqBrack} precisely align with the behavior of inhomogeneous Teukolsky amplitudes under rotation. This consequence of how Eqs.~\eqref{eq:phiPhase} and \eqref{eq:rPhase} constrain the evolving frequencies reflects the orbital symmetries that relate rotations ($\phi_p^{(j)}\rightarrow \phi_p^{(j+1)}$) to time-translations ($t^{(j)}\rightarrow t^{(j+1)}$).

According to Eq.~\eqref{eq:periPhase}, orbital phases (like $\phi_p$ and $v$) are sufficient for determining the waveform phases at each periastron, but additional considerations are required between periastrons. By interpreting $(j)$ to represent the most recent periastron, the orbital phases can be used to approximate the waveform phases at any time through combination of Eqs.~\eqref{eq:phase_int} and \eqref{eq:periPhase}
\begin{align}
\label{eq:phaseRecent}
&\Phi_{mn}(t) = \Phi_{mn}^{(j)} + m \int_{t^{(j)}}^t \Omega_\phi \, dt + n \int_{t^{(j)}}^t \Omega_r \, dt.
\end{align}
Because $(j)$ represents the most recent periastron, these integrals involve rather short intervals ($t-t^{(j)} \sim M$). Unlike long duration integrals (with intervals consisting of the entire inspiral), it is a good approximation to consider the slowly evolving frequencies as constants in Eq.~\eqref{eq:phaseRecent} (this strategy was adopted in~\cite{WarbOsbu17}):
\begin{align}
\Phi_{mn}(t) = m\, \phi^{(j)}_p + n\, v^{(j)} + (m\Omega_\phi + n &\Omega_r)(t-t^{(j)}) \notag
\\& + \mathcal{O}(\epsilon).
\label{eq:phaseOld}
\end{align}
Although this representation of the waveform phase is not exact, Eq.~\eqref{eq:phaseOld} satisfies our overall phase accuracy requirements. In applying Eq.~\eqref{eq:phaseOld}, it is essential to refresh the frequencies and orbital phases, $\Omega_\phi$, $\Omega_r$, $\phi^{(j)}_p$, $v^{(j)}$, and $t^{(j)}$, with updated values (new constants) after each new periastron or else the constant frequency approximation will deteriorate as $t-t^{(j)}$ grows larger.

Because Eq.~\eqref{eq:phaseOld} involves the original (slow to compute) orbital variables, it would be convenient to formulate the waveform phases in terms of the (fast to compute) NIT orbital variables. One important ingredient in Eq.~\eqref{eq:phaseOld} is the fundamental frequencies, which can be approximated using the NIT equations of motion
\begin{align}
\label{eq:approxFreq}
&\Omega_r = \frac{d\tilde{v}}{d\tilde{t}} + \mathcal{O}(\epsilon) , \qquad\;\; \Omega_\phi = \frac{d\tilde{\phi}_p}{d\tilde{t}}  + \mathcal{O}(\epsilon) .
\end{align}
However, recent work suggests that Eq.~\eqref{eq:approxFreq} is one order more accurate than we claim here~\cite{Lynch}. Another vital ingredient in Eq.~\eqref{eq:phaseOld} is the orbital phases evaluated at perisatrons (those with $(j)$ superscripts). Thankfully, the accumulated orbital phases at periastron are largely interchangeable (within $\mathcal{O}(\epsilon)$ tolerance) with their NIT equivalents at periastron (approximate agreement at periastron without needing to invert is a feature of this NIT application)
\begin{align}
\Phi_{mn}(t) = m\, \tilde{\phi}^{(j)}_p + n\, \tilde{v}^{(j)} + (m\Omega_\phi  + n & \Omega_r)(t-\tilde{t}^{(j)}) \notag
\\ & + \mathcal{O}(\epsilon).
\label{eq:phaseJump}
\end{align}
The explicit appearance of $t$ in Eq.~\eqref{eq:phaseJump} suggests that the inverse NIT must be applied to $\tilde{t}$ (but not $\tilde{\phi}_p$) to generate time-domain Teukolsky waveforms from NIT inspirals (kludge waveforms based on NIT inspirals would need to invert both $\tilde{t}$ and $\tilde{\phi}_p$ in applying Eq.~\eqref{eq:kludge}).

\begin{figure}
\includegraphics[width=3.3in]{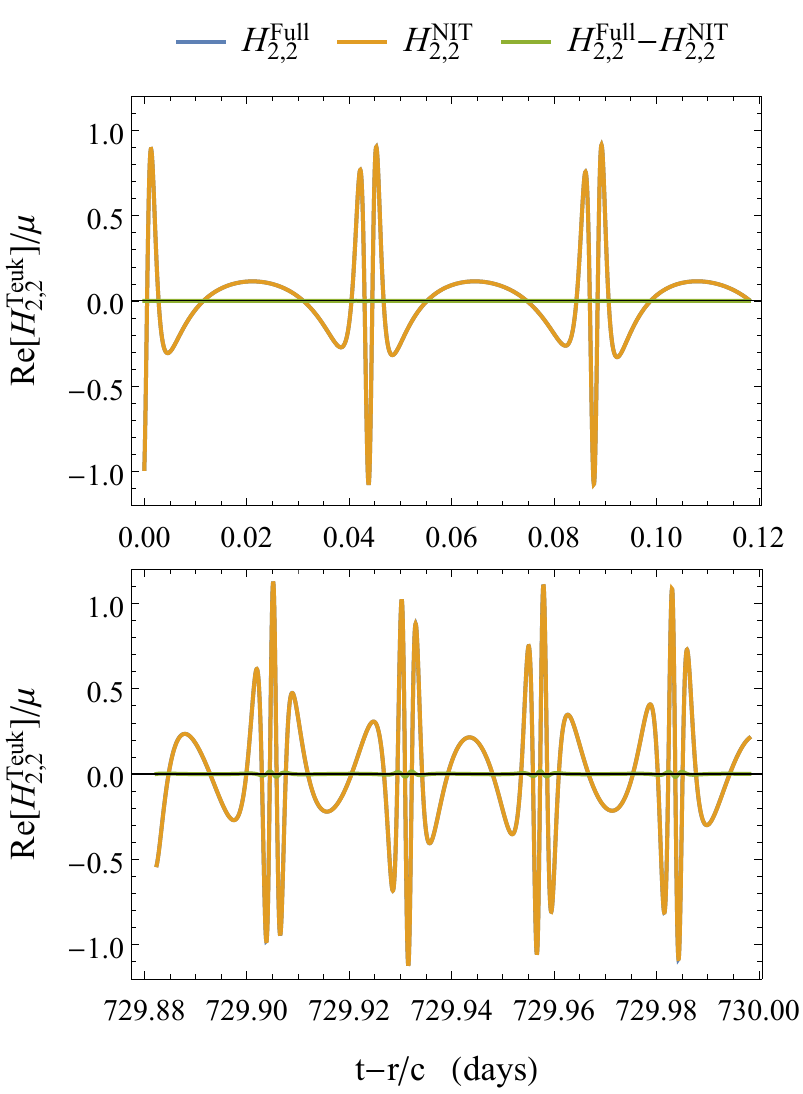}
\caption{\label{fig:FullVsNIT} Comparison of two strategies for post-processing self-forced inspirals to calculate relativistic Teukolsky waveforms. This plot demonstrates that our streamlined waveform generation method based on NIT parameters (Eq.~\eqref{eq:Hfinal}) is reliable. The blue and yellow waveforms have the same extreme mass-ratio ($\epsilon=10^{-5}$ with $M=10^6 M_\odot$) and initial orbital configuration ($p_0=11$, $e_0=0.7$). The top frame shows the early (matched) waveforms while the bottom frame shows independently evolved (blue and orange) waveforms shortly before plunge. The blue curve is based on the ``true" (but slow to compute) orbital phases from the full self-force inspiral. That scheme involves discontinuously advancing orbital phases after each periastron and calculating phase growth between periastrons by assuming constant frequencies during each radial period~\cite{WarbOsbu17}. The orange curve is based on the fast-to-compute but approximate NIT orbital phases in a new scheme (Eq.~\eqref{eq:Hfinal}) where the waveform phase is recovered through careful linear combination of smooth NIT orbital phases. The full self-force inspiral consumed $\sim 173$~seconds of computing time while the NIT method consumed only $\sim 0.0401$~seconds of computing time ($\sim 4000$ times faster); however, inverting $\tilde{t}$ and re-sampling $t$ consumed an additional $\sim 1.27$ seconds (perhaps future algorithmic enhancements are possible, see Sec.~\ref{sec:CandFD}). Notice that the two methods overlap perfectly enough that the blue curve is not visible behind the orange curve. The green curve displays the absolute error. This close alignment between waveform calculations from NIT parameters and those involving the ``true" orbital phases validates our new streamlined NIT-based waveform model. Incidentally, for at least the present configuration this favorable comparison also validates the underlying NIT inspiral, but there are other configurations (larger $\epsilon$) where the NIT technique itself is the limiting factor (see Sec.~\ref{sec:FullVsNIT}).}
\end{figure}

As it stands, Eq.~\eqref{eq:phaseJump} involves unfortunate discontinuities during the transition from one orbital period to the next. These small but sudden jumps in the waveform phases are consequences of resetting the orbital phases after each new periastron (to balance the constant frequency approximation of Eq.~\eqref{eq:phaseRecent}). For EMRIs, these $\mathcal{O}(\epsilon)$ discontinuities are largely negligible. However, such artifacts may be enhanced late in the inspiral (near breakdown of adiabaticity) and/or in the case of IMRIs. Furthermore, there is potential for artificial discontinuities to be magnified during Fourier transform (a key data analysis technique, see Sec.~\ref{sec:overlap}). Although it is debatable whether this property of Eq.~\eqref{eq:phaseJump} would adversely affect LISA data analysis in practice, there is a remedy in any case. This remedy follows from a qualitative observation regarding accumulation of smooth NIT orbital phases. While the original motivation for counting periastrons was to enforce Eqs.~\eqref{eq:phiPhase} and \eqref{eq:rPhase} as constraints, those $(j)$ superscripts can be qualitatively interpreted (in the context of Eq.~\eqref{eq:phaseJump}) to represent average phase accumulation at a slowly changing rate. Unlike the standard orbital parametrization, the NIT orbital phases already accumulate with rates that inherently evolve slowly (even between periastrons). Therefore, we argue that, within $\mathcal{O}(\epsilon)$ tolerance, the $(j)$ superscripts in Eq.~\eqref{eq:phaseJump} (and their exclusive periastron interpretation) can be carefully abolished for accumulated NIT phases
\begin{align}
\Phi_{mn}(t) = m\, \tilde{\phi}_p  + n\, \tilde{v} + (m\Omega_\phi + n \Omega_r)(&t-\tilde{t}) \notag
\\&+ \mathcal{O}(\epsilon) .
\label{eq:phaseSmooth}
\end{align}
The quantitative justification supporting Eq.~\eqref{eq:phaseSmooth} is derived from the following short-duration properties of the approximate frequencies described in Eq.~\eqref{eq:approxFreq}
\begin{align}
\tilde{\phi}_p - \tilde{\phi}^{(j)}_p  =  \int_{\chi^{(j)}}^{\chi} \left(\frac{d\tilde{\phi}_p}{d\tilde{t}}\right)\left(\frac{d\tilde{t}}{d\chi}\right) d\chi = \, &\Omega_\phi \left(\tilde{t} - \tilde{t}^{(j)}\right) \notag
\\& + \mathcal{O}(\epsilon) ,
\\
\tilde{v} - \tilde{v}^{(j)}  =  \int_{\chi^{(j)}}^{\chi} \left(\frac{d\tilde{v}}{d\tilde{t}}\right)\left(\frac{d\tilde{t}}{d\chi}\right) d\chi = \Omega_r & \left(\tilde{t} - \tilde{t}^{(j)}\right) \notag
\\&  + \mathcal{O}(\epsilon) ,
\end{align}
which is sufficient to demonstrate that Eqs.~\eqref{eq:phaseJump} and \eqref{eq:phaseSmooth} are consistent up to $\mathcal{O}(\epsilon)$. By also making the replacement $(p,e) \rightarrow (\tilde{p},\tilde{e})$ in determining $A_{lmn}$ (they differ only by $\mathcal{O}(\epsilon)$) we arrive at our representation of the relativistic Teukolsky waveform in terms of NIT parameters
\begin{align}
\label{eq:Hfinal}
H^\text{Teuk}_{lm}(t) = \sum_n A_{lmn}(\tilde{p},\tilde{e}) \; &e^{-i \,[m\, \tilde{\phi}_p  + n\, \tilde{v}  + (m\Omega_\phi + n \Omega_r)(t-\tilde{t})]}.
\end{align}
Besides the need to calculate $t$ by inverting $\tilde{t}$ (a notably time-intensive process~\cite{VandWarb18}), Eq.~\eqref{eq:Hfinal} smoothly and efficiently outputs relativistic Teukolsky waveforms from NIT inspiral data while avoiding additional integrals for the waveform phases (guaranteeing consistency with orbital phases), which is a novel result. Figure~\ref{fig:FullVsNIT} demonstrates consistency between waveforms generated from (fast to compute) NIT inspiral parameters and those generated from (slow to compute)  full self-force inspiral parameters (note that both methods use fully relativistic Teukolsky amplitudes). As an aside, in applications that combine the NIT equations of motion with a two-timescale expansion of the field equations (such as calculating the second-order self-force for eccentric orbits), Eq.~\eqref{eq:phaseSmooth} could be a candidate to represent the phase for each mode of the metric perturbation. Convenient two-timescale related properties include phase ingredients ($\tilde{\phi}_p$, $\tilde{v}$, and $\tilde{t}$) with fast-time derivatives that, when combined, should depend only on slow-time.

\subsection{Measuring overlap of competing templates}
\label{sec:overlap}

Much of our discussion has revolved around competing techniques where one is more reliable (involves fewer approximations) while the other is more practical (faster, less complicated, etc.). One competition involves the slow but reliable full self-force inspirals vs. the fast but approximate NIT inspirals. Another competition involves the complicated but reliable evolving Teukolsky snapshots (multivoice) vs. the simple but approximate kludge waveform amplitudes. It is important to evaluate whether a more practical technique maintains consistency with the associated reliable technique. If consistency were achieved it would validate deployment of templates based on the more practical technique (actualizing associated LISA data analysis enhancements). Or perhaps consistency is achievable in only some regions of parameter space while in other areas there could be limitations that stem from a breakdown of approximations (with a quantifiable breakdown threshold). In any case, we seek an appropriate method to quantify consistency between competing waveform templates (focusing particularly on the high eccentricities that are achievable through this work).

We measure the level of agreement between two waveforms by adapting a standard gravitational wave data analysis technique (see~\cite{Babak_2007,Cutler_1994,Finn_1992} for detailed discussion). This technique quantifies the overlap of two waveforms by defining an appropriate inner product, $\langle x|y \rangle$, that accounts for detector sensitivity as a function of frequency
\begin{align}
    \langle x|y \rangle = \max_{\Delta t} \left[ \int_{-\infty}^{+\infty} \frac{X^{*}\,Y\,e^{-2\pi i f \Delta t} + Y^{*}\,X\,e^{2\pi i f \Delta t}}{S(f)} df \right]  . \label{eq:overlapEqn}
\end{align}
Here $x(t)$ and $y(t)$ represent two time domain waveforms while $X(f)$ and $Y(f)$ are their Fourier transforms.  The complex exponentials in the integrand account for the possibility of an arbitrary time translation, $\Delta t$. We compute all Fourier transforms ($x\rightarrow X$ and $y\rightarrow Y$) discretely using the fftw3 library~\cite{FFTW05}. Similarly, we use the discrete Fourier transform and the inverse discrete Fourier transform as appropriate to evaluate the overlap integral for a discrete set of possible time translations (whichever $\Delta t$ value causes the largest overlap is adopted, as implied by the ``$\max$" operation in Eq.~\eqref{eq:overlapEqn}). For LISA's noise power spectral density, $S(f)$, we adopt an appropriate expression from~\cite{Robson_2019}. We are particularly interested in the ``fractional overlap", which is normalized so that the overlap score has a maximum size of $1$, 
\begin{align}
    \text{Fractional Overlap} = \frac{\langle x|y \rangle}{\sqrt{\langle x|x \rangle \langle y|y \rangle}} \, . \label{eq:overlapEqnNormalized}
\end{align}
When the fractional overlap score is less than $\sim 0.95$ it is likely to adversely affect parameter estimation during LISA data analysis~\cite{Babak_2007}, although stricter criteria may be necessary depending on context. As an application of Eq. \ref{eq:overlapEqnNormalized}, consider the overlap score of the two waveforms presented in Fig.~\ref{fig:FullVsNIT}, which have a fractional overlap of $0.999701$.

\section{Results}
\label{sec:results}

\subsection{Full self-force vs. NIT inspirals}
\label{sec:FullVsNIT}

\begin{figure}
  \includegraphics[width=3.35in]{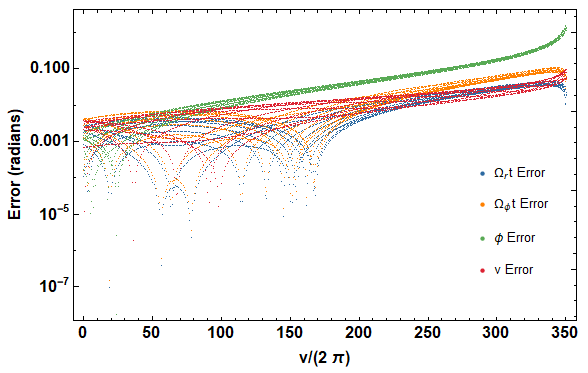} 
  \caption{Errors in the NIT inspiral parameters ($\Omega_r t, \Omega_{\phi}t, \phi_p, v$) vs. the accumulated number of radial oscillations are shown for $\epsilon = 10^{-3}$ with $p_0 = 11$ and $e_0 = 0.7$. The error is found via the absolute difference between the full self-force inspiral value (a reliable benchmark) and the NIT value (after applying the inverse NIT). Note that this figure captures dephasing of the NIT and full self-force inspirals with a maximum error in $\phi_p$ of $\sim 1.4$ radians (after a total inspiral duration of only $\sim 2.1$ days with $M=10^6 M_\odot$). Although the NIT error naturally grows rapidly near plunge as adiabaticity deteriorates, the error had already grown to $\sim$0.41 radians at 95\% of total radial oscillations.} 
  \label{fig3}
\end{figure}

In this section the main comparisons and evaluations of competing models are conducted. First we compare inspiral models to determine when the fast NIT method produces inspirals and waveforms that are consistent with the reliable (but slow) full self-force method. Here we adopt equivalent waveform generation schemes (based on Teukolsky amplitudes) during these inspiral evaluations, but we will turn our attention to competing waveform schemes in Sec.~\ref{sec:TeukVsKludge}. Prior to examining waveform overlaps directly, we compare orbital predictions from NIT and full self-force inspirals because orbital discrepancies should foreshadow waveform discrepancies. The accumulation of orbital phase error over time for a high eccentricity IMRI ($e_0=0.7$, $\epsilon=10^{-3}$) is shown in Fig.~\ref{fig3}. The large accumulation of error in $\phi_p$ ($\sim 1.4$ radians for a total duration of only $\sim 2.1$~days with $M=10^6 M_\odot$) suggests that the fast NIT method does not maintain consistency with the full self-force benchmark for these particular parameters (intermediate $\epsilon$ with large $e_0$). However, Fig.~\ref{fig:FullVsNIT} previously demonstrated that the NIT method was highly successful for a more extreme mass-ratio ($\epsilon = 10^{-5}$) with the same $e_0$ and $p_0$. The existence of an $\epsilon$ dependent threshold describing the range of validity of the NIT technique is perhaps not surprising considering that the method follows from truncating an expansion in powers of $\epsilon$  (as seen in Eqs.~\ref{eq:NITi}-\ref{eq:NITf}). Our truncation after 1\textsuperscript{st} post-adiabatic order suggests an NIT phase error (after inversion) of size $\epsilon$, which we hypothesize is responsible for the large error illustrated in Fig.~\ref{fig3}.

\begin{figure}
\centering
  \vspace{0.2cm}
  \includegraphics[width=3.35in]{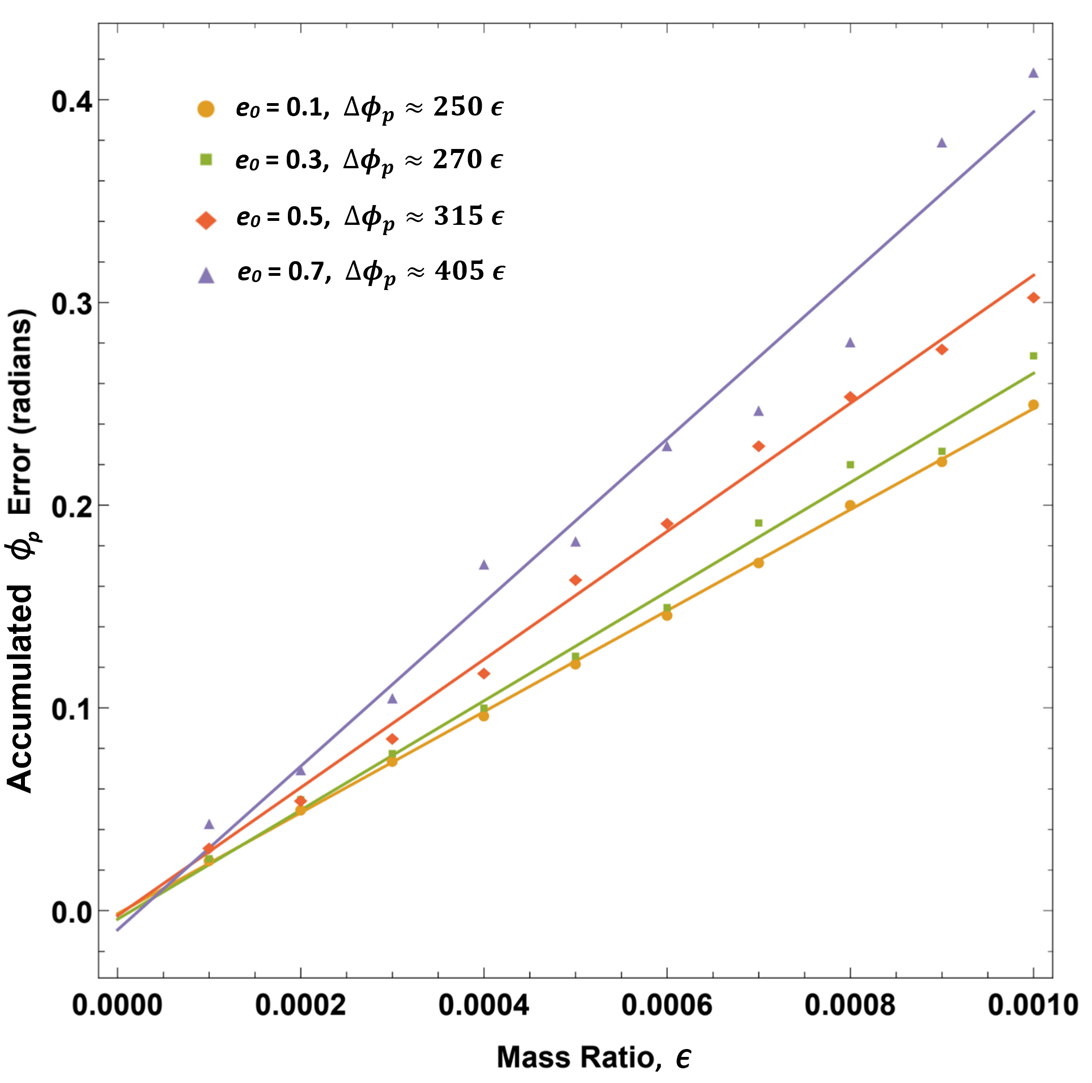} 
    \caption{Accumulated NIT $\phi_p$ error is plotted vs. mass-ratio for various initial eccentricities and fixed $p_0 = 11$. To assess the accuracy of $\phi_p$ calculated from the inverse NIT, we use full self-force inspirals with equivalent initial parameters as a reliable benchmark. Note that the largest error comes from the data point corresponding to Fig.~\ref{fig3} ($\epsilon = 0.001$, $e_0 = 0.7$, $p_0 = 11$, $M=10^6 M_\odot$). The total inspiral duration for that largest error is only $\sim 2.1$ days (so it can be worse for longer durations). Although the trend is roughly linear in $\epsilon$ (implying that the missing 2\textsuperscript{nd} post-adiabatic correction is to blame), these data involve considerable noise. That noise is a consequence of how the computation suddenly and unpredictably crashes near plunge. From a numerical standpoint, when $p-2e\rightarrow 6$ the denominator in certain equations of motion approaches zero. Will that vanishing denominator cause a crash during this radial oscillation, or the next, or the one after that? The associated unpredictabilities introduce tens of radians of uncertainty at plunge. We scale down those uncertainties by extracting error values after only 95\% of total radial oscillations have been completed (still noisy because that 95\% is measured relative to crashing, but steady enough to observe a linear trend).}
  \label{fig9}
\end{figure}

To test that hypothesis we have compared NIT and full self-force inspirals for many $e_0$ and $\epsilon$ values (at fixed $p_0=11$). Figure \ref{fig9} demonstrates a linearly increasing relationship between the NIT phase error and the mass-ratio, which is consistent with our hypothesis that missing 2\textsuperscript{nd} post-adiabatic terms with large coefficients are responsible for NIT dephasing at intermediate mass-ratios. There is also an increasing relationship between the initial eccentricity and the NIT phase error. We decided to investigate whether these discrepancies could involve implementation flaws rather than limitations of the NIT technique itself. After doubling and quadrupling the number of frequencies in our Fourier decomposition of the self-force and doubling the resolution of our orbital parameter interpolation, we observed the same NIT breakdown for highly eccentric IMRIs. Besides, such implementation flaws would cause an $\mathcal{O}(\epsilon^0)$ error rather than the observed $\mathcal{O}(\epsilon^1)$ error. To further verify these observations, we next analyze the accuracy of waveforms based on NIT inspirals under conditions relevant to LISA. 

\begin{figure}
\centering
  \vspace{0.2cm}
  \includegraphics[width=3.2in]{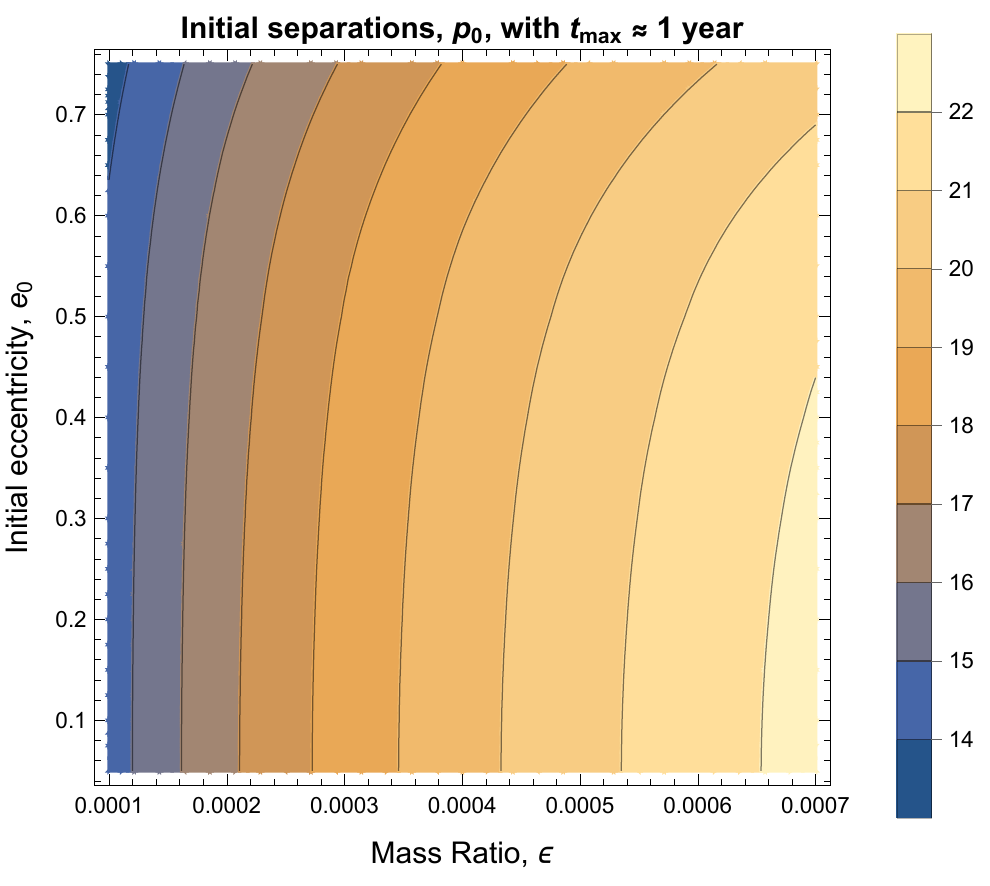} 
    \caption{The $p_0$ values (as a function of $e_0$ and $\epsilon$) that fix the inspiral duration at $t_\text{max}\simeq 1$~year are displayed (each contour involves a constant $p_0$). The primary mass is fixed at $M=10^6 M_\odot$. Larger $p_0$ values are needed to maintain $t_\text{max}$ when $\epsilon$ increases. Increasing $e_0$ allows for slightly smaller $p_0$ values at fixed $t_\text{max}$. These $t_\text{max}\simeq 1$~year combinations specify the desired subset of our full self-force vs. NIT waveform overlap data.}
  \label{fig10}
\end{figure}

\begin{figure}
\centering
  \vspace{0.2cm}
  \includegraphics[width=3.35in]{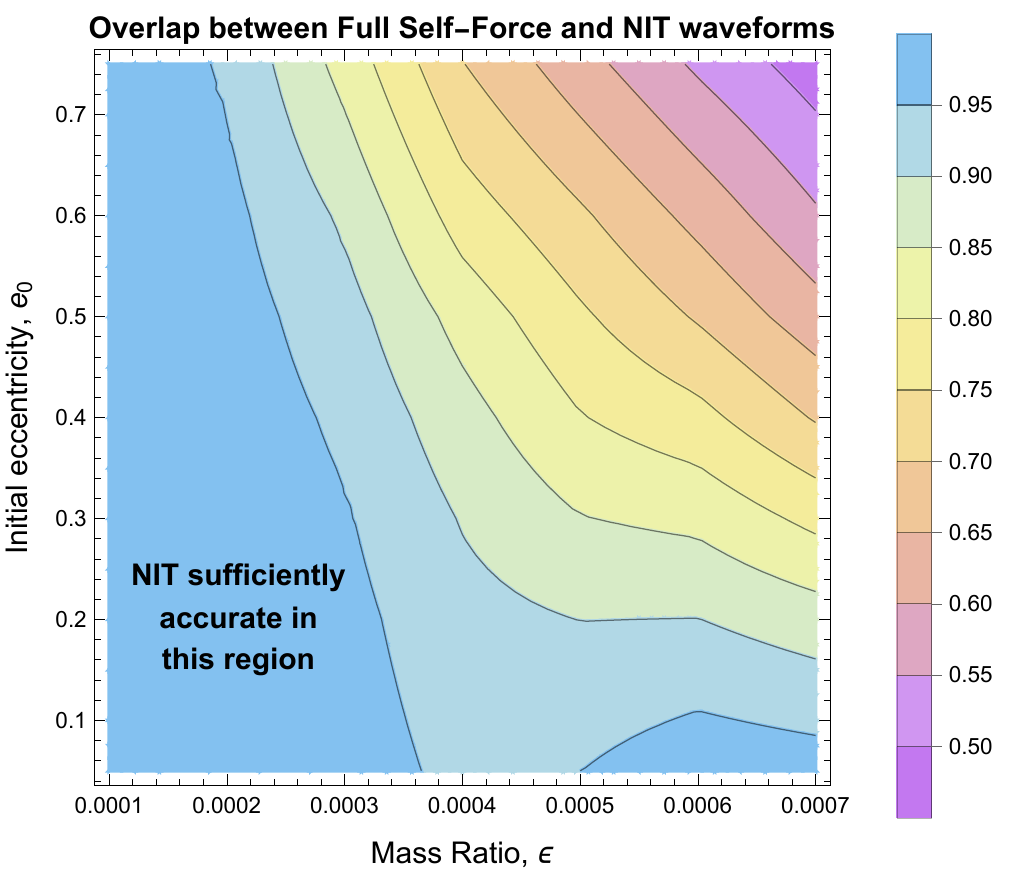} 
    \caption{Fractional overlap (as a function of $e_0$ and $\epsilon$) of full self-force and NIT waveforms are displayed (each contour involves a constant fractional overlap). All templates under comparison have a fixed waveform duration of $t_\text{max}\simeq 1$~year (see Fig.~\ref{fig10}). The waveforms are based on ($l$,$m$)~=~(2,2) Teukolsky amplitudes (phases are inferred from orbital properties, see Sec.~\ref{sec:TeukWave}). The NIT method (through 1\textsuperscript{st} post-adiabatic order) deteriorates when $e_0$ and $\epsilon$ are large. Possible improvements are discussed in Sec.~\ref{sec:CandFD}.} 
  \label{fig11}
\end{figure}

One important template property relevant to LISA observations is the waveform duration. Accordingly, we seek waveforms based on inspirals with a total time interval spanning a significant fraction of LISA's mission duration. To provide a concise account of the accuracy attainable by relevant NIT inspiral parameters, we calculate appropriate fractional waveform overlaps as a function of $e_0$ and $\epsilon$ while fixing the total inspiral duration at $\simeq 1$~year. To achieve that outcome, we first calculate Teukolsky waveforms for both NIT and full self-force inspirals (carefully measuring each fractional overlap) with the following combinations of parameters: $p_0 = \{13,14,...,23\}$, $e_0 = \{0.1,0.2,...,0.7\}$, $\epsilon = \{0.0001,0.0002,...,0.0007\}$. Each of these waveforms have their own total duration, $t_\text{max}$. By interpolating $p_0$ as a function of $t_\text{max}$, $e_0$, and $\epsilon$, we determined the $p_0$ for each $e_0$ and $\epsilon$ that achieves $t_\text{max}\simeq 1$~year (see Fig.~\ref{fig10}). These carefully determined $p_0$ values, along with their associated $e_0$ and $\epsilon$, reveal the subset of our overlap data corresponding to $t_\text{max}\simeq 1$~year (we have essentially reduced the dimensionality of parameter space by constraining $p_0$ to fix $t_\text{max}$). Those fractional overlaps are displayed as a function of $e_0$ and $\epsilon$ in Fig.~\ref{fig11}. Notice that the fractional overlap scores indicate inadequate agreement between the NIT and full self-force model waveforms for large $\epsilon$ and $e_0$, which implies deterioration of inverse NIT phases in that region of parameter space; this measurement of eccentricity dependent NIT deterioration is a new result. Largely, lower eccentricity NIT inspirals are able to maintain accuracy deeper into the intermediate mass-ratio regime while the NIT method may require further enhancement (2\textsuperscript{nd} post-adiabatic corrections) to accurately model high eccentricity IMRI waveforms. Beyond confirming the qualitative features of Fig.~\ref{fig9}, Fig.~\ref{fig11} illustrates the limitations of our current NIT implementation under conditions more closely related to LISA data analysis.

\subsection{Teukolsky vs. kludge waveforms}
\label{sec:TeukVsKludge}

Having quantified the range of validity for NIT phase approximations, we similarly seek to determine when waveform amplitudes derived from kludges are consistent with reliable Teukolsky models. While past comparisons have considered temporary snapshots~\cite{Babak_2007}, here we consider time intervals involving entire inspirals (with associated evolution of orbital properties). For each comparison we calculate a self-forced worldline via the NIT method and post-process that self-forced worldline to independently generate a kludge waveform and a Teukolsky waveform. Some kludge implementations also introduce their own rough approximation of the worldline, but it is well known that self-forced inspirals are necessary to achieve phase accuracy requirements (so we do not consider kludged worldlines here). Note that usage of identical worldlines for the competing waveforms prevents dephasing, which focuses our analysis on the accuracy of synthesized (summing over harmonics) waveform amplitudes associated with weak-field kludge approximations.

\begin{figure}
\centering
  \vspace{0.2cm}
  \includegraphics[width=3.35in]{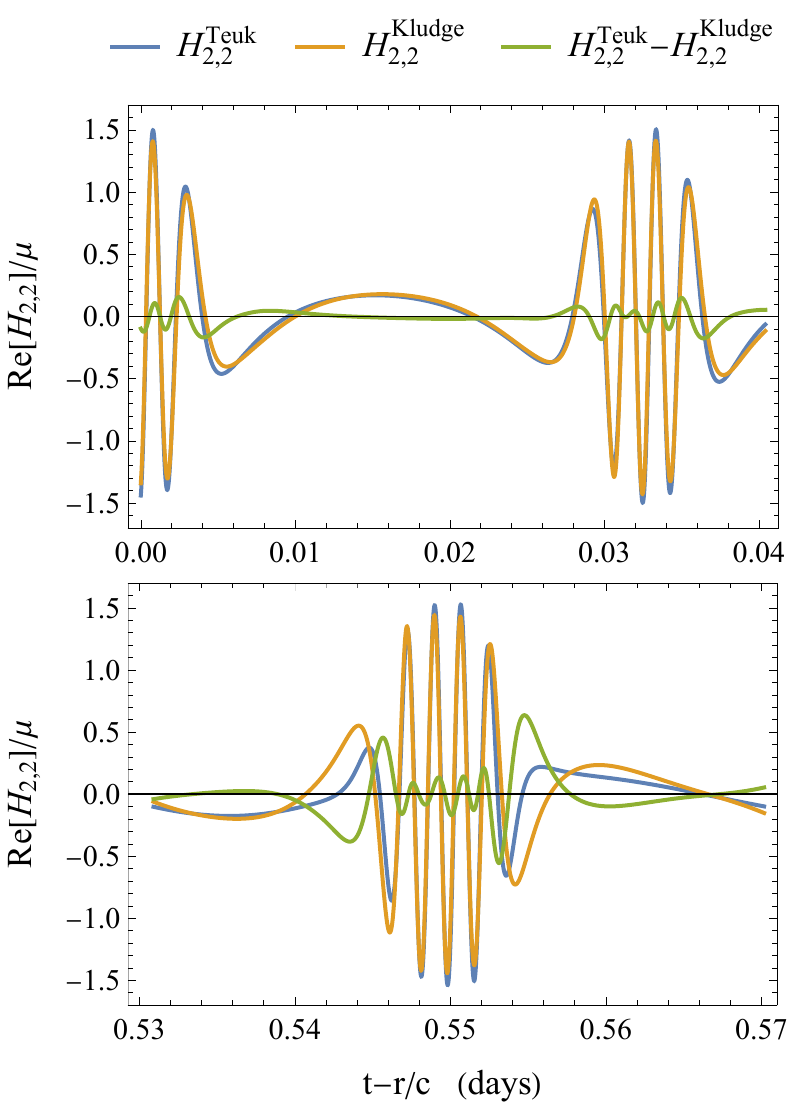} 
   \caption{A Teukolsky waveform and kludge waveform calculated from identical NIT inspiral worldlines are depicted. The associated parameters are $\epsilon = 2\times 10^{-4}$, $M = 10^6 M_\odot$, $p_0 = 7.75$, and $e_0 = 0.7$. We have considered such a small initial separation because the weak-field kludge approximation is most likely to fail when $p$ is small. Notice that the absolute error of the kludge waveform is small at early times but becomes larger late in the inspiral. Overall, the fractional overlap of these two waveforms is 0.9897. Perhaps this seemingly strong overlap score is a consequence of the near-plunge kludge breakdown occurring during only a small fraction of total cycles. We have uniformly applied a relative time translation of 7 seconds between the two waveforms to maximize their overlap. Generally, we found that increasing the sampling rate to 1 Hz was sufficient to resolve the optimal time translation (in the context of Eq.~\eqref{eq:overlapEqn}) at high eccentricities.}
  \label{fig:TvsW}
\end{figure}

\begin{figure}
\centering
  \vspace{0.2cm}
  \includegraphics[width=3.35in]{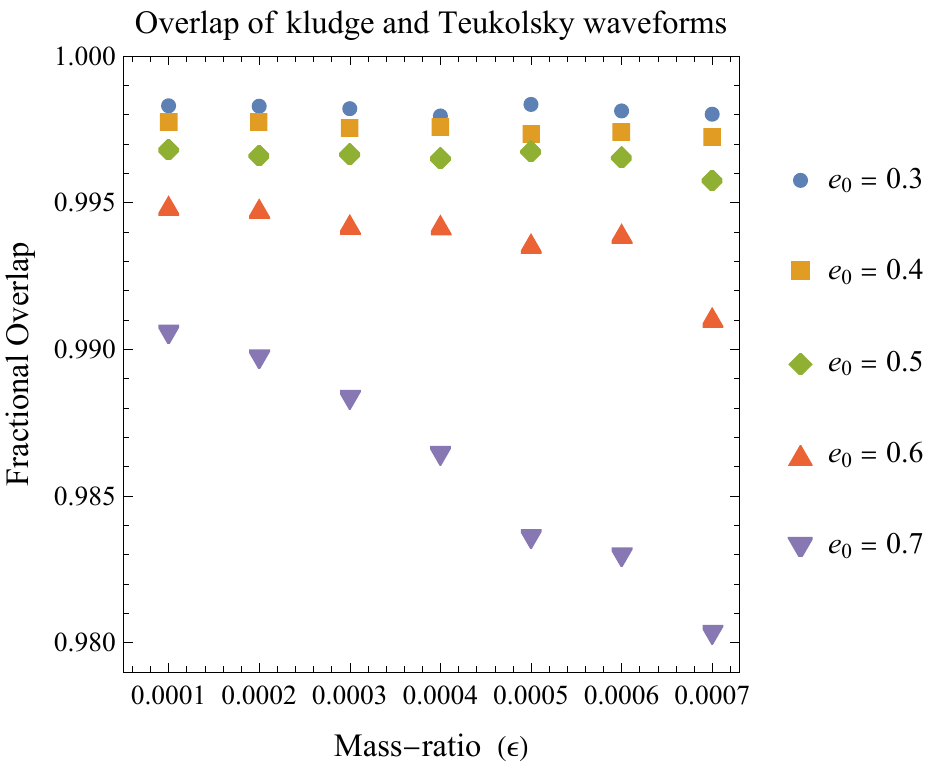} 
   \caption{Fractional overlap scores comparing Teukolsky and kludge waveforms are shown for various initial eccentricities. Each inspiral and waveform fixes the following parameters: $M = 10^6 M_\odot$ and $p_0 = 7.75$. These parameter choices focus on the strong-field regime to investigate potential breakdown of the kludge model. For this non-spinning case, the kludge model performs well with overlaps greater than $\sim 0.98$. Generally we observe that the kludge waveform accuracy decreases with increasing $e_0$ or $\epsilon$. The noisy variations in the data are likely related to how the inspiral crashes somewhat unpredictably near plunge (see Fig.~\ref{fig9} for more details).}
  \label{fig:tw_overlap}
\end{figure}

Figure~\ref{fig:TvsW} shows a kludge waveform and Teukolsky waveform calculated from identical NIT inspiral worldlines. Although the absolute error of the kludge model grows when $p$ becomes small, that breakdown occurs during only a small fraction of total cycles (which may explain the seemingly strong fractional overlap of 0.9897). Similar comparisons for various initial eccentricities are depicted in Fig.~\ref{fig:tw_overlap}. For this non-spinning case, the kludge model performs well with overlaps greater than $\sim 0.98$, which confirms that phase accuracy outweighs amplitude accuracy even when summing over harmonics associated with eccentric motion. Generally, we observe that the kludge waveform accuracy decreases with increasing $e_0$.

\section{Conclusions and Future Directions \label{sec:CandFD}}

By interfacing our highly eccentric self-force library with the Fast Self-forced Inspirals package~\cite{VandWarb18} of the Black Hole Perturbation Toolkit~\cite{toolkit}, we have quantified how higher eccentricities affect various features (accuracy, speed, etc.) of inspiral and waveform models. Our analysis comprised of two investigations that each probe a different aspect of competing techniques. One investigation compared slow (debilitatingly slow) but reliable full self-force inspirals to fast but approximate NIT inspirals. Our NIT implementation is capable of calculating highly eccentric $\sim 2$~year EMRI worldlines with only $\sim 40$~ms of computing time ($\sim 4000$ times faster than the full self-force method). In evaluating whether the fast NIT inspirals are capable of achieving necessary phase accuracy goals at high eccentricity, we determined that our implementation is sufficiently accurate for EMRIs ($\epsilon \lesssim 2\times 10^{-4}$, see Fig.~\ref{fig11}). However, we have identified certain large absent 2\textsuperscript{nd}~post-adiabatic coefficients (see Fig.~\ref{fig9}) whose omission inhibits NIT accuracy for highly eccentric IMRIs ($\epsilon \gtrsim 2\times 10^{-4}$). To illustrate this challenge, consider an enhanced transformation between $\phi_p$ and $\tilde{\phi}_p$ (elevating Eq.~\eqref{eq:phitilde} to 2\textsuperscript{nd}~post-adiabatic order)
\begin{align}
    \tilde{\phi}_p &= \phi_p + Z^{(0)}_\phi(p,e,v) + \epsilon \, Z^{(1)}_\phi(p,e,v) + \mathcal{O}(\epsilon^2)\, .
\end{align}
Hypothetically, if $Z^{(1)}_\phi$ ever became sufficiently large (like for high eccentricity), it could explain the observed NIT error for IMRIs (Fig.~\ref{fig9} might imply $Z^{(1)}_\phi\sim 10^{2} - 10^{3}$). Although, in practice one would need to include all 2\textsuperscript{nd}~post-adiabatic terms in the NIT expansion (beyond $Z^{(1)}_\phi$ alone) to comprehensively improve the expansion order. Upon recognizing one area where including 2\textsuperscript{nd}~post-adiabatic terms would provide benefits, we now speculate that other areas may similarly benefit from calculations beyond 1\textsuperscript{st}~post-adiabatic order. One example is 3\textsuperscript{rd}~order self-force calculations, which would be needed if their associated 2\textsuperscript{nd} post-adiabatic coefficient ever becomes large (it is not known how large that coefficient is, but we have shown that large 2\textsuperscript{nd} post-adiabatic coefficients exist elsewhere in highly eccentric IMRI models). Note that these uniquely high eccentricity limitations are not in conflict with recent evidence~\cite{warburton2021gravitationalwave,Rifat_2020} suggesting that quasi-circular models based on black hole perturbation theory maintain a level of validity in the comparable mass limit (notice in Fig.~\ref{fig11} how the accurate region extends towards large $\epsilon$ when $e_0$ is small). 

Another investigation compared synthesized (summing over harmonics) waveform amplitudes based on a weak-field kludge approximation to reliable (but less accessible) Teukolsky amplitudes. To interface the NIT method with evolving Teukolsky snapshots, we developed a novel representation of the waveform phases based on NIT inspiral parameters (see Eq.~\eqref{eq:phaseSmooth}). Besides the need to invert $\tilde{t}$, our technique efficiently and smoothly produces Teukolsky waveforms while avoiding additional integrals for the waveform phases in a way that guarantees consistency with orbital phases. In comparing kludge waveforms to Teukolsky waveforms (for identical self-forced worldlines), we found that the weak-field kludge approximation deteriorates with increasing eccentricity. However, for this non-spinning case the kludge amplitude model performs well (this is not true for kludge inspiral models) with fractional overlaps greater than $\sim 0.98$ (the late-time kludge amplitude breakdown is only a small fraction of the total inspiral duration). Kludge amplitude models are more likely to fail in more extreme scenarios such as prograde inspirals into a rapidly rotating black hole (which venture deeper into the strong-field regime); this should justify further development of Teukolsky models. 

Future directions beyond those mentioned above could involve more comprehensive self-force calculations (fully post-adiabatic), generalization to more extensive astronomical scenarios, and/or optimizations during waveform template production. Including the fully post-adiabatic self-force would require treatment of the non-geodesic history during 1\textsuperscript{st} order self-force calculations and incorporation of 2\textsuperscript{nd}~order self-force results~\cite{PounETC20,warburton2021gravitationalwave}. Both of these missing elements are accessible through a two-timescale expansion~\cite{HindFlan08,Miller_2021,pound2021black}. Generalization to more extensive astronomical scenarios suggests consideration of spin for the smaller body ~\cite{BurkKhan15,Ruangsri_2016,WarbOsbu17,Piovano_2020}, the larger body ~\cite{WarbBara11,Vand16,Vand18,NasiOsbu19,nasipak2021resonant,Lynch}, or both. Optimizing production of waveform templates could involve incorporating highly eccentric self-force results (as with this work) into existing techniques like those featuring machine learning + GPU acceleration~\cite{Chua_2021,katz2021fastemriwaveforms} and/or waveform assembly directly in the frequency domain~\cite{Hughes_2021} (our Eq.~\eqref{eq:Hfinal} seems to be a perfect fit for a direct frequency domain approach).

\acknowledgements

J. M. and J. B. gratefully ackowledge support from the New York Space Grant. T. O. gratefully acknowledges support from the SUNY Geneseo Presidential Fellowship. We thank Alvin Chua, Scott Hughes, Philip Lynch, Adam Pound, Maarten van de Meent, Niels Warburton, and the anonymous referee for helpful comments. This work makes use of the Black Hole Perturbation Toolkit~\cite{VandWarb18}.

\appendix

\section{Kludge amplitude derivation example}
\label{sec:kludge_app}

As an example we demonstrate a rough derivation of $H^\text{kludge}_{2,2}$ (see Chapter~3.3 of~\cite{maggiore} for full details leading to Eq.~\eqref{eq:endmag}). The well-known quadruple formula approximates the far-zone spatial metric from weak-field sources
\begin{align}
\label{eq:quadrupoleformula}
&h_{ij}^\text{quad}=\frac{2}{r}\ddot{I}_{ij} \, ,
\end{align}
where $I_{ij}$ is the quadrupole moment tensor. Treating the small body as a point pass, the quadrupole moment tensor takes the form $I_{ij}=\mu \, x^i_p x^j_p$, where $x^i_p$ is the small body's position. The Cartesian components of the quadrupole moment tensor are
\begin{align}
\label{eq:qmtValues}
&I_{ij}=
\left(
\begin{array}{ccc}
 \mu \, r_p^2 \, \cos ^2\phi_p & \mu \, r_p^2 \, \sin\phi_p \,\cos \phi_p & 0 \\
 \mu \, r_p^2 \, \sin\phi_p \, \cos\phi_p & \mu \, r_p^2 \, \sin^2\phi_p & 0 \\
 0 & 0 & 0 \\
\end{array}
\right)
\end{align}
It is convenient to convert to spherical coordinates ($x^{k'}$)
\begin{align}
\label{eq:convert}
&h^\text{quad}_{k'n'} = \frac{\partial x^i}{\partial x^{k'}}\frac{\partial x^j}{\partial x^{n'}}\, h^\text{quad}_{ij} .
\end{align}
The two gravitational wave polarizations are found from the transverse metric components
\begin{align}
\label{eq:hplusfull}
h^\text{quad}_+ &=\frac{4}{r^3}h^\text{quad}_{\theta \theta} \notag
\\&= \frac{4\mu}{r}\cos^2\theta\,\Big(2\cos^2(\phi-\phi_p)\dot{r}_p^2
\\&\;\;\;\; +4r_p\sin(2\phi-2\phi_p)\dot{r}_p\dot{\phi}_p+2r_p\cos^2(\phi-\phi_p)\ddot{r}_p\notag
\\&\;\;\;\; -2r_p^2\cos(2\phi-2\phi_p)\dot{\phi}_p^2+r_p^2\sin(2\phi-2\phi_p)\ddot{\phi}_p\Big) \, , \notag
\\
h^\text{quad}_\times&= \frac{4}{r^3}\sin{\theta}\,h^\text{quad}_{\theta \phi} \notag
\\&= \frac{2\mu}{r}\sin\theta\,\sin(2\theta)\,\Big(-\sin(2\phi-2\phi_p)\dot{r}_p^2
\\&\;\;\;\; +4r_p\cos(2\phi-2\phi_p)\dot{r}_p\dot{\phi}_p-r_p\sin(2\phi-2\phi_p)\ddot{r}_p\notag
\\&\;\;\;\; +2r_p^{2}\sin(2\phi-2\phi_p)\dot{\phi}_p^2+r_p^2\cos(2\phi-2\phi_p)\ddot{\phi}_p\Big) \, . \notag \label{eq:endmag}
\end{align}
The quadrupole results can be expressed as a sum over $s=-2$ spin-weighted spherical harmonics with $l=2$
\begin{align}
&h^\text{quad}_+ - i\, h^\text{quad}_\times =   \frac{1}{r} \Big( H^\text{kludge}_{2,-2} \,_{-2}Y_{2,-2}(\theta,\phi) 
\\&\qquad\qquad + H^\text{kludge}_{2,0} \,_{-2}Y_{2,0}(\theta,\phi) + H^\text{kludge}_{2,2} \,_{-2}Y_{2,2}(\theta,\phi)  \Big) . \notag
\end{align}
To find $H^\text{kludge}_{2,2}$ (the example term we provided) we leverage the orthogonality of the spin-weighted spherical harmonics
\begin{align}
\label{eq:isolateKludge}
&H_{2,2}^\text{kludge}&=\int r(h^\text{quad}_+ - i\, h^\text{quad}_\times)\;_{-2}Y_{2,2}^{*}(\theta, \phi)\;d\Omega\, ,
\end{align}
which produces Eq.~\eqref{eq:kludge}.

To apply this result in our scheme, we must express it in terms of our orbital parameters: $p$, $e$, and $v$ ($\phi_p$ already appears explicitly). Equation~\eqref{eq:dar} provides $r_p$ (and so $\frac{dr_p}{dv}$), but we also need $\frac{d\phi_p}{dv}$ and $\frac{dv}{dt}$ (which are available from tangent geodesic properties) to find $\dot{r}_p$ and $\dot{\phi}_p$
\begin{align}
    &\frac{dr_p}{dv} = \frac{e\,p\,M\,\sin{v}}{(1+e\,\cos{v})^2} ,
    \\
    &\frac{d\phi_p}{dv} = \sqrt{\frac{p}{p-6-2 e\,\cos{v} }}
    \\
    &\frac{dv}{dt} = \frac{M(p-2-2e\,\cos{v})}{r_p^2} \sqrt{\frac{p-6-2 e\,\cos{v} }{(p-2)^2-4e^2}}
    \\
    &\dot{r}_p = \left(\frac{dr_p}{dv}\right) \left({\frac{dv}{dt}}\right) , \\
    &\dot{\phi}_p = \left(\frac{d\phi_p}{dv}\right) \left({\frac{dv}{dt}}\right) .
\end{align}
Similar logic provides $\ddot{r}_p$ and $\ddot{\phi}_p$ (we assume $\dot{p}$ and $\dot{e}$ are negligible because they are suppressed by a factor of $\epsilon$).
Finally, $H_{2,2}^\text{kludge}$ can be expressed in terms of $v$, $p$, $e$, and $\phi_p$
\begin{align}
\label{eq:isolateKludge}
&H_{2,2}^\text{kludge}=-\mu \frac{p-2-2 e \cos{v}}{p^2 \big((p-2)^2-4 e^2\big)}\sqrt{\frac{\pi }{5}} e^{-2i\phi_p} \Bigg( 8 p^2  \notag
\\&  -2 e^2 (4 e^2-2 p^2+5 p+30) \cos (2 v) -3 e^4 \cos (4 v) \notag
\\&+e\big(e^2 (p-56)+4 (3 p^2-4 p-12)\big) \cos{v} -5 e^4 \notag
\\&-e^3 p \cos (3 v) -24 e^3 \cos (3 v) +10 e^2 p -52 e^2 -16 p \notag
\\&+2ie \sqrt{p(p-6-2 e \cos{v})} \Big( 2 e p \sin (2 v) - e^2 \sin{v}\notag
\\&-e^2 \sin (3 v)-4 e \sin (2 v)+4 p \sin{v}-4 \sin{v} \Big) \Bigg).
\end{align}

\bibliography{main}

\end{document}